\documentclass{article}
\usepackage{spconf,amsmath,graphicx}
\usepackage{comment}
\usepackage{cite}
\usepackage[utf8]{inputenc}
\usepackage{xcolor}
\usepackage{url}
\usepackage{amssymb}
\usepackage{subfigure}

\usepackage[linesnumbered,ruled,vlined,commentsnumbered]{algorithm2e}

\SetCommentSty{mycommfont}

\renewcommand{\v}[1]{\ensuremath{\mathbf{#1}}}
\DeclareMathOperator*{\argmax}{arg\,max}

\newcommand{\pdffigure}[3][width=0.7\linewidth]{
	\begin{figure}[tb]
	\begin{center}
    \IfFileExists{./#2.pdf}{
	    \includegraphics[#1]{#2.pdf}
	}{
	    \includegraphics[draft]{#2.pdf}
    }
    \end{center}
    \caption{#3}
	\label{fig:#2}
	\end{figure}
}

\title{End-to-end neural speaker diarization with self-attention}
\name{
Yusuke Fujita$^{1,2}$
\thanks{The first author performed the work while at Center for Language and Speech Processing, Johns Hopkins University as a Visiting Scholar.}\!\!,
Naoyuki Kanda$^{1}$\!,
Shota Horiguchi$^{1}$\!,
Yawen Xue$^{1}$\!,
Kenji Nagamatsu$^{1}$\!, 
Shinji Watanabe$^{2}$}
\address{
$^{1}$ \text{Hitachi, Ltd. Research \& Development Group, Japan} \\
$^{2}$ \text{Center for Language and Speech Processing, Johns Hopkins University, USA}}
\begin{document}
\ninept
\maketitle
\begin{abstract}
Speaker diarization has been mainly developed based on the clustering of speaker embeddings.
%The dominant approach to speaker diarization has been clustering of speaker embeddings. 
However, the clustering-based approach has two major problems; i.e., (i) it is not optimized to minimize diarization errors directly, and (ii) it cannot handle speaker overlaps correctly.
To solve these problems, the End-to-End Neural Diarization (EEND), in which a bidirectional long short-term memory (BLSTM) network directly outputs speaker diarization results given a multi-talker recording, was recently proposed.
In this study, we enhance EEND by introducing self-attention blocks instead of BLSTM blocks.
In contrast to BLSTM, which is conditioned only on its previous and next hidden states, self-attention is directly conditioned on all the other frames, making it much suitable for dealing with the speaker diarization problem.
We evaluated our proposed method on simulated mixtures, real telephone calls, and real dialogue recordings.
The experimental results revealed that the self-attention was the key to achieving good performance and that our proposed method performed significantly better than the conventional BLSTM-based method. Our method was even better than that of the state-of-the-art x-vector clustering-based method.
%, while it correctly handled speaker overlaps.
%Finally, by visualizing the latent representation, we show that the similarity between distant frames is actually captured by the self-attention blocks.
Finally, by visualizing the latent representation, we show that the self-attention can capture global speaker characteristics in addition to local speech activity dynamics.
Our source code is available online at \url{https://github.com/hitachi-speech/EEND}.

\end{abstract}
\section{Introduction}
\label{sec:intro}
Speaker diarization is the process of partitioning an audio recording into homogeneous segments according to the speaker's identity. The speaker diarization has a wide range of applications, such as information retrieval from broadcast news, generating minutes of meetings, and a turn-taking analysis of telephone conversations \cite{Tranter2006, Anguera2012}.
It also helps automatic speech recognition performance in multi-speaker conversation scenarios in meetings (ICSI \cite{Janin03, etin2006OverlapIM}, AMI \cite{Renals2008, Kanda2019ICASSP}) and home environments (CHiME-5 \cite{Barker2018, Du2018, Boeddecker2018, Kanda2018, Kanda2019ICASSP}).

Typical speaker diarization systems are based on the clustering of speaker embeddings \cite{Meignier2010LIUMSA,Shum2013,Sell2014, Senoussaoui2014,Dimitriadis2017, Romero2017, Maciejewski2018CharacterizingPO, Wang2018LSTM}.
For instance, i-vectors \cite{Dehak2011, Shum2013, Sell2014, Maciejewski2018CharacterizingPO}, d-vectors\cite{Wan2018, Wang2018LSTM}, and x-vectors \cite{Snyder2018, Romero2017} are commonly used in speaker diarization tasks.
These embeddings of short segments are partitioned into speaker clusters by using clustering algorithms, such as Gaussian mixture models \cite{Meignier2010LIUMSA, Shum2013}, agglomerative hierarchical clustering \cite{Meignier2010LIUMSA, Sell2014, Romero2017, Maciejewski2018CharacterizingPO},
mean shift clustering \cite{Senoussaoui2014}, k-means clustering \cite{Dimitriadis2017, Wang2018LSTM}, Links \cite{Mansfield2018,Wang2018LSTM}, and spectral clustering \cite{Wang2018LSTM}.
These clustering-based diarization methods have shown themselves to be effective on various datasets (see the DIHARD Challenge 2018 activities, e.g.,  \cite{Sell2018dihard,Diez2018,Sun2018}).

However, such clustering-based methods have a number of problems.
First, they cannot be optimized to minimize diarization errors directly, because the clustering procedure is a type of unsupervised learning methods.
Second, they have trouble handling speaker overlaps, since the clustering algorithms implicitly assume one speaker per segment.
Furthermore, they have trouble adapting their speaker embedding models to real audio recordings with speaker overlaps, because the speaker embedding model has to be optimized with single-speaker non-overlapping segments.
These problems hinder the speaker diarization application from working on real audio recordings that usually contain overlapping segments.

To solve these problems, we propose Self-Attentitive End-to-End Neural Diarization (SA-EEND).
Different from most of the other methods, our proposed method does not rely on clustering.
Instead, a self-attention-based neural network directly outputs the joint speech activities of all speakers for each time frame, given an input of a multi-speaker audio recording.
Our method can naturally handle speaker overlaps during the training and inference time by exploiting a multi-label classification framework.
The neural network is trained in an end-to-end fashion using a recently proposed permutation-free objective function that provides minimal diarization errors \cite{Fujita2019E2EDiarization}.

This paper shows that our method achieves a significant performance improvement over end-to-end neural diarization (EEND) \cite{Fujita2019E2EDiarization}, for which promising but preliminary results were reported with a bidirectional long short-term memory (BLSTM) \cite{Graves2005}. 
In particular, it shows that the self-attention mechanism \cite{Lin2017,Vaswani2017} is the key to achieving good speaker-diarization performance in this paper.
We demonstrate that the self-attention mechanism gives significantly better results for multiple datasets compared with the BLSTM-based method \cite{Fujita2019E2EDiarization} and the state-of-the-art x-vector-based speaker diarization method. 
In contrast to BLSTM, which is conditioned only on its previous and next hidden states, the self-attention layer is conditioned on all the other input frames by computing the pairwise similarity between all frame pairs.
We believe that this mechanism is the key to speaker diarization since it can capture global speaker characteristics in addition to local speech activity dynamics.
By visualizing the learned representation, we show that some self-attention heads capture speaker-dependent global characteristics, while the remaining heads represent temporal features.
\section{Related work}
\label{sec:relatedwork}

\subsection{Clustering-based methods}

\begin{figure}[tb]
  \begin{center}
  \subfigure[X-vector clustering-based method]{
    \includegraphics[width=0.95\linewidth]{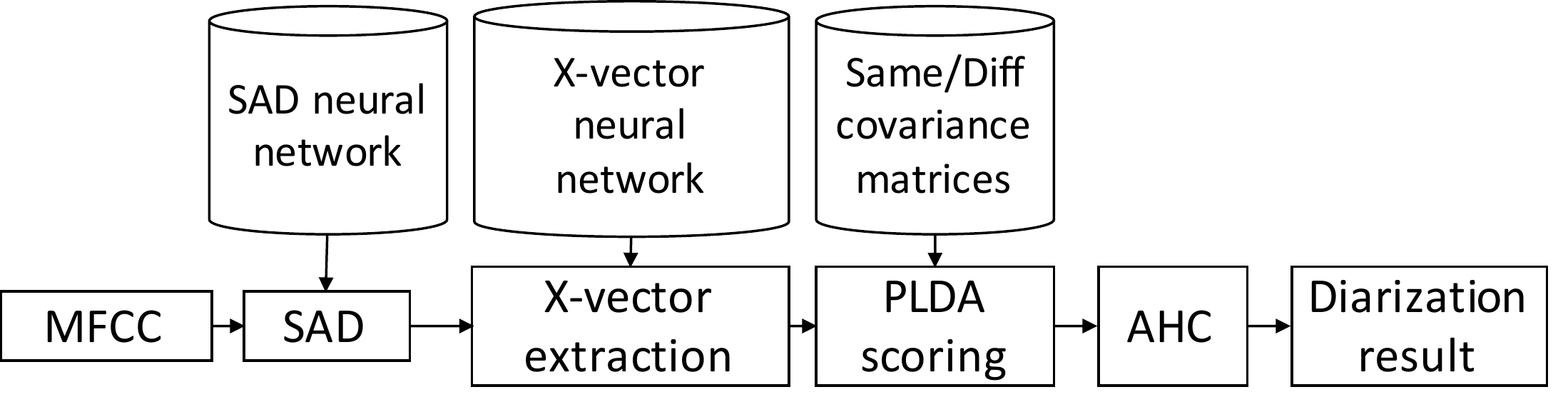}
    \label{fig:x-vector}
  }
  %\hfill
  \subfigure[EEND method]{
    \includegraphics[width=0.95\linewidth]{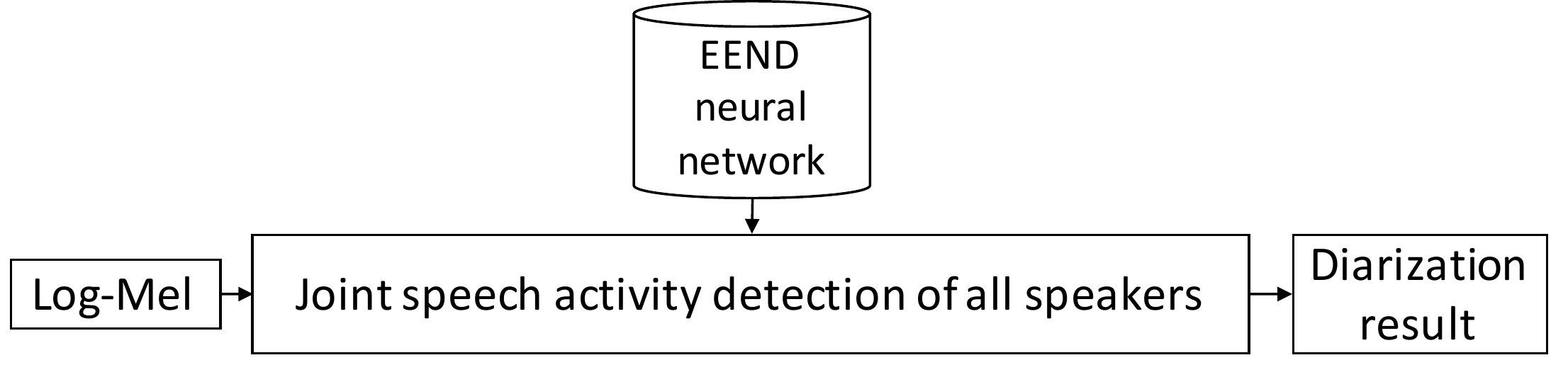}
    \label{fig:eend}
  }
  \end{center}
  \vspace{-0.5cm}
  \caption{System diagrams for speaker diarization}
  \label{fig:systems}
\end{figure}

The x-vector clustering-based system is commonly used for speaker diarization \cite{Sell2018dihard, Diez2018, Snyder2019}.
A diagram of the system is depicted in Fig. \ref{fig:x-vector}.
To build the system, one has to prepare three independent models: (i) a speech activity detection (SAD) neural network, (ii) x-vector extraction neural network, and (iii) PLDA model including the same/different speaker covariance matrices.
None of these models can be trained to directly minimize the diarization errors.
Joint modeling methods have been studied in an effort to alleviate the complex preparation process and take into account the dependencies between these models.
They include, for example, joint modeling of x-vector extraction and PLDA scoring \cite{Romero2017, Narayanaswamy2019} and joint modeling of SAD and speaker embedding \cite{MiasatoFilho2018}.
However, the clustering process has remained unchanged because it is an unsupervised process.

In contrast to these methods, the EEND method uses only one neural network model, as depicted in Fig. \ref{fig:eend}. 
This method does not rely on clustering, and the model can be directly optimized with the reference diarization results of the training data.

This neural-network-based end-to-end approach, in which only one neural network model directly computes the final outputs, has been successfully applied in a variety of tasks, including neural machine translation \cite{Bahdanau2015, Sutskever2014}, automatic speech recognition \cite{Chorowski2015, Chan2016, Watanabe2017}, and text-to-speech \cite{Wang2017,Sotelo2017}.

\subsection{Direct optimization minimizing diarization errors}
A fully supervised diarization method has been proposed for optimization based on a diarization error minimization objective \cite{Zhang2018}. 
This is the first successful approach that does not cluster speaker embeddings. 
The method formulates the speaker diarization problem on the basis of a factored probabilistic model, which consists of modules for determining speaker changes, speaker assignments, and feature generation.
These models are jointly trained using input features and corresponding speaker labels. However, the SAD model and their speaker embedding (d-vector) model have to be trained separately in their method. Moreover, their speaker-change model assumes one speaker for each segment, which hinders its application to speaker-overlapping speech.

In contrast to their method, the EEND method uses an end-to-end neural network that accepts audio features as input and outputs the joint speech activities of multiple speakers. The network is optimized using the entire recording, including non-speech and speaker overlaps, with a diarization-error-oriented objective. This end-to-end model was first introduced in \cite{Fujita2019E2EDiarization}; this paper describes an extension of the model that includes a self-attention mechanism.

\subsection{Self-attention mechanism}

The self-attention mechanism was originally proposed for extracting sentence embeddings for text processing \cite{Lin2017}. Recently, the self-attention mechanism has shown superior performance in a variety of tasks, including machine translation \cite{Vaswani2017}, video classification \cite{Wang2018NonlocalNN}, and image segmentation \cite{Ye_2019_CVPR}. For audio processing, a self-attention mechanism has been incorporated in acoustic modeling for ASR \cite{Sperber2018, Dong2018}, sound event detection \cite{Wang2018}, and speaker recognition \cite{Zhu2018}. For speaker diarization, the self-attention mechanism has been applied to the speaker embedding extraction model \cite{Sun2018} and the scoring model \cite{Narayanaswamy2019} of clustering-based methods. This study describes a self-attention mechanism for clustering-free speaker diarization.

\section{Proposed method: Self-Attentive End-to-End Neural Diarization}
\label{sec:method}

\subsection{End-to-end neural diarization: review}
\label{sec:e2e}
Here, we describe the EEND method proposed in \cite{Fujita2019E2EDiarization}.
The speaker diarization task can be formulated as a multi-label classification problem, as follows.

Given a $T$-length observation sequence $X = (\v{x}_t \in \mathbb{R}^{F} \mid t=1,\cdots,T)$ from an audio signal, speaker diaization problem tries to estimate the corresponding speaker label sequence $Y = (\v{y}_t \mid t=1,\cdots,T)$.
Here, $\v{x}_t$ is a $F$-dimensional observation feature vector at time index $t$.
Speaker label $\v{y}_t = [y_{t,c} \in \{0,1\} \mid c=1, \cdots, C]$ denotes a joint activity for multiple ($C$) speakers at time index $t$.
For example, $y_{t,c} = 1$ and $y_{t,c'} = 1\; (c \ne c')$ represent an overlap situation in which speakers $c$ and $c'$ are both present at time index $t$.
Thus, determining $Y$ is a sufficient condition to determine the speaker diarization information.

The most probable speaker label sequence $\hat{Y}$ is selected from among all possible speaker label sequences $\mathcal{Y}$, as follows:
\begin{equation}
\hat{Y} = \argmax_{Y \in \mathcal{Y}} P(Y|X).
\end{equation}
$P(Y|X)$ can be factorized using the conditional independence assumption as follows:
\begin{align}
    P(Y|X) &= \prod_{t} P(\v{y}_t | \v{y}_1, \cdots \v{y}_{t-1}, X), \\
    &\approx \prod_t P(\v{y}_t| X) \approx \prod_t \prod_c P(y_{t,c}| X).
\end{align}
Here, we assume that the frame-wise posterior is conditioned
on all inputs, and each speaker is present independently.
The frame-wise posterior $P(y_{t,c}| X)$ can be estimated using a neural-network-based model.

\subsection{Self-attention-based neural network}

In \cite{Fujita2019E2EDiarization}, a BLSTM based neural network was used for estimating the frame-wise posteriors $P(y_{t,c}| X)$.
In this paper, we propose self-attentive end-to-end neural diarization (SA-EEND), which uses self-attention-based encoding blocks instead of BLSTMs, as depicted in Fig. \ref{fig:self-att-pit}.
The input features are transformed as follows:
\begin{align}
    \v{e}^{(0)}_t &= \mathbf{W}_0\v{x}_t + \mathbf{b}_0 \in \mathbb{R}^{D}, \\
    \v{e}^{(p)}_t &= \mathrm{Encoder}^{(p)}_t(\v{e}^{(p-1)}_1, \cdots, \v{e}^{(p-1)}_T) \  (1 \le  p \le P). \label{eq:hidden}
\end{align}
Here,  $\mathbf{W}_0 \in \mathbb{R}^{D \times F}$ and $\mathbf{b}_0 \in \mathbb{R}^D $ project an input feature into $D$-dimensional vector.
$\mathrm{Encoder}^{(p)}_t(\cdot)$ is the $p$-th encoder block which accepts an input sequence of $D$-dimensional vectors and outputs a $D$-dimensional vector $\v{e}_t^{(p)}$ at time index $t$. We use $P$ encoder blocks followed by the output layer for frame-wise posteriors.

The architecture of the encoder block is depicted in Fig. \ref{fig:self-att-pit}.
This configuration of the encoder block is almost the same as the one in the Speech-Transformer introduced in \cite{Dong2018}, but without positional encoding.
The encoder block has two sub-layers. The first is a multi-head self-attention layer, and the second is a position-wise feed-forward layer.
\subsubsection{Multi-head self-attention layer}
The multi-head self-attention layer transforms a sequence of input vectors as follows.
The sequence of vectors $(\mathbf{e}^{(p-1)}_t | t = 1, \cdots, T)$ is converted into a $\mathbb{R}^{T \times D}$ matrix, followed by layer normalization \cite{Ba2016}:
\begin{equation}
    \mathbf{\bar{E}}^\mathrm{(p-1)} = \mathrm{LayerNorm}([\mathbf{e}^{(p-1)}_1 \cdots \mathbf{e}^{(p-1)}_T]^\top) \in \mathbb{R}^{T \times D}. \label{eq:ln1}
\end{equation}
Then, for each head, a pairwise similarity matrix $\mathbf{A}^{(p)}_h$ is computed using the dot products of query vectors $\mathbf{\bar{E}}^{(p-1)} \mathbf{Q}^{(p)}_h \in \mathbb{R}^{T \times d}$ and key vectors $\mathbf{\bar{E}}^{(p-1)} \mathbf{K}^{(p)}_h \in \mathbb{R}^{T \times d}$:
\begin{equation}
        \mathbf{A}^{(p)}_h = \mathbf{\bar{E}}^{(p-1)} \mathbf{Q}^{(p)}_h (\mathbf{\bar{E}}^{(p-1)}  \mathbf{K}^{(p)}_h)^\top \in \mathbb{R}^{T \times T} \: (1\le h \le H), \label{eq:a}
\end{equation}
where, $\mathbf{Q}^{(p)}_h, \mathbf{K}^{(p)}_h \in \mathbb{R}^{D \times d}$ are query and key projection matrices for the $h$-th head, respectively.
$d = D/H$ is a dimension of each head, and $H$ is the number of heads.
The pairwise similarity matrix $\mathbf{A}^{(p)}_h$ is scaled by $1/\sqrt{d}$ and a softmax function is applied to form the attention weight matrix $\mathbf{\hat{A}}^{(p)}_h$:
\begin{equation}
        \mathbf{\hat{A}}^{(p)}_h = \mathrm{Softmax}\left(\frac{\mathbf{A}^{(p)}_h}{ \sqrt{d}}\right) \in \mathbb{R}^{T \times T}. \label{eq:softmax}
\end{equation}
Then, using the attention weight matrix, context vectors $\mathbf{C}^{(p)}_h$ are computed as a weighted sum of the value vectors $\mathbf{\bar{E}}^{(p-1)} \mathbf{V}_h^{(p)} \in \mathbb{R}^{T \times d}$:
\begin{equation}
        \mathbf{C}^{(p)}_h = \mathbf{\hat{A}}^{(p)}_h (\mathbf{\bar{E}}^{(p-1)}\mathbf{V}_h^{(p)}) \in \mathbb{R}^{T \times d}, \label{eq:value}
\end{equation}
where $\mathbf{V}_h \in \mathbb{R}^{D \times d}$ is the value projection matrix.
Finally, the context vectors for all heads are concatenated and projected using the output projection matrix $\mathbf{O}^{(p)} \in \mathbb{R}^{D \times D}$:
\begin{equation}
    \mathbf{E}^{(p,\mathrm{SA})} = [\mathbf{C}^{(p)}_1 \cdots \mathbf{C}^{(p)}_H] \mathbf{O}^{(p)} \in \mathbb{R}^{T \times D}. \label{eq:concat}
\end{equation}
Following the self-attention layer, a residual connection and layer normalization is applied:
\begin{equation}
    \mathbf{\bar{E}}^{(p,\mathrm{SA})} = \mathrm{LayerNorm}(\mathbf{\bar{E}}^{(p-1)} + \mathbf{E}^{(p,\mathrm{SA})}) \in \mathbb{R}^{T \times D}. \label{eq:res}
\end{equation}

\subsubsection{Position-wise feed-forward layer}

The position-wise feed-forward layer transforms $\mathbf{\bar{E}}^{(p,\mathrm{SA})}$ as follows:
\begin{equation}
    \mathbf{E}^{(p,\mathrm{FF})} = \mathrm{ReLU}( \mathbf{\bar{E}}^{(p,\mathrm{SA})}\mathbf{W}^{(p)}_1 + \mathbf{b}^{(p)}_1 \mathbf{1}) \mathbf{W}^{(p)}_2 + \mathbf{b}^{(p)}_2 \mathbf{1} \in \mathbb{R}^{T \times D},
\end{equation}
where $\mathbf{W}^{(p)}_1 \in \mathbb{R}^{D \times d_\mathrm{ff}}$ and $\mathbf{b}^{(p)}_1 \in \mathbb{R}^{d_\mathrm{ff}}$ are the first linear projection matrix and bias, respectively, $\mathbf{1} \in \mathbb{R}^{1 \times T}$ is an all-one row vector, and $\mathrm{ReLU}(\cdot)$ is the rectified linear unit activation function.
$d_\mathrm{ff}$ is the number of internal units in this layer. 
$\mathbf{W}^{(p)}_2 \in \mathbb{R}^{d_\mathrm{ff} \times D}$ and $\mathbf{b}^{(p)}_2 \in \mathbb{R}^{D}$ are the second linear projection matrix and bias, respectively.

Finally, the output of the encoder block $\mathbf{e}_t^{(p)}$ for each time frame is computed by applying a residual connection as follows:
\begin{equation}
    [\mathbf{e}_1^{(p)} \cdots \mathbf{e}_T^{(p)}] = (\mathbf{\bar{E}}^{(p,\mathrm{SA})} + \mathbf{E}^{(p,\mathrm{FF})})^\top
\end{equation}

\subsubsection{Output layer for frame-wise posteriors}

The frame-wise posteriors $\mathbf{z}_t$ are calculated from $\mathbf{e}_t^{(P)}$ (in Eq. \ref{eq:hidden}) using layer normalization and a fully-connected layer as follows:
\begin{align}
\bar{\mathbf{E}}^{(P)}&=\mathrm{LayerNorm}([\mathbf{e}_1^{(P)} \cdots \mathbf{e}_T^{(P)}]^\top) \in \mathbb{R}^{T\times D}, \\
[\mathbf{z}_1 \cdots \mathbf{z}_T] &= \sigma(\bar{\mathbf{E}}^{(P)} \mathbf{W}_3 + \mathbf{b}_3 \mathbf{1})^\top, \label{eq:out}
\end{align}
where $\mathbf{W}_3 \in \mathbb{R}^{D \times C}$ and $\mathbf{b}_3 \in \mathbb{R}^{C}$ are the linear projection matrix and bias, respectively, and $\sigma(\cdot)$ is the element-wise sigmoid function.

\subsection{Permutation-free training}

The difficulty of training the model described above is that the model must deal with speaker permutations: changing the order of speakers within a correct label sequence is also regarded as correct. An example of permutations in a two-speaker case is shown in Fig. \ref{fig:self-att-pit}.
In this paper, we call this problem ``label ambiguity.'' This label ambiguity obstructs the training of the neural network when we use a standard binary cross entropy loss function.

\pdffigure[width=\linewidth]{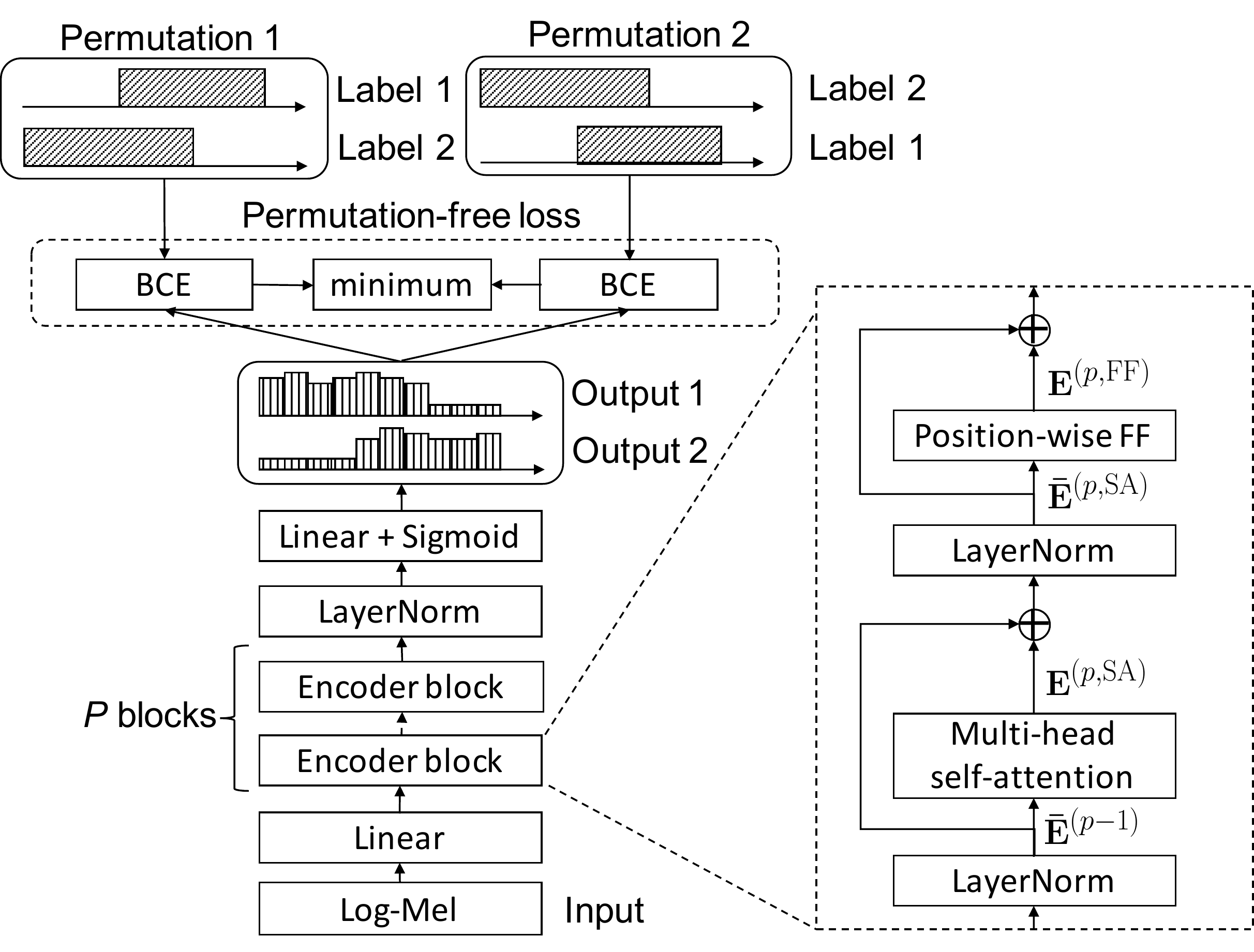}{Two-speaker SA-EEND model trained with permutation-free loss.}

To cope with the label ambiguity problem, the permutation-free training scheme considers all the permutations of the reference speaker labels.
The permutation-free training scheme has been used in research on source separation \cite{Hershey2016, Yu2017, Kolbak2017}. Here, we apply the permutation-free loss function to a temporal sequence of speaker labels.
The neural network is trained to minimize the permutation-free loss between the output $\v{z}_t$ predicted in Eq. \ref{eq:out} and the reference speaker label $\v{l}_t$, as follows:
\begin{equation}\label{eq:pf}
    J^{\text{PF}} = \frac{1}{TC} \min_{\phi \in \mathrm{perm}(C)} \sum_t \mathrm{BCE}(\v{l}_t^\phi, \v{z}_t),
\end{equation}
where $\mathrm{perm}(C)$ is the set of all the possible permutations of ($1,\dots,C$),
and $\v{l}_t^\phi$ is the $\phi$-th permutation of the reference speaker label, and $\mathrm{BCE}(\cdot, \cdot)$ is the binary cross entropy function between the label and the output.

\section{Experimental setup}
\label{sec:setup}

\subsection{Data}
\label{sec:data}

\begin{table}[t]
\caption{Statistics of training and test sets.}
\label{tab:set}
\centering
\begin{tabular}{l|rrr} \hline
    & \# mixtures & avg. duration & \multicolumn{1}{c}{overlap} \\
    & & (sec) & ratio (\%) \\\hline
Traning sets & & & \\
\: Simulated ($\beta=2$) & 100,000 & 87.6 & 34.4 \\
\: Real (SWBD+SRE) & 26,172 & 304.7 & 3.7 \\ \hline
Test sets & & & \\
\: Simulated ($\beta=2$) & 500 & 87.3 & 34.4 \\
\: Simulated ($\beta=3$) & 500 & 103.8 & 27.2 \\
\: Simulated ($\beta=5$) & 500 & 137.1 & 19.5 \\
\: CALLHOME \cite{callhome} & 148 & 72.1 & 13.0 \\
\: CSJ \cite{Maekawa2003} & 54 & 766.3 & 20.1 \\ \hline
\end{tabular}
\end{table}

To verify the effectiveness of the SA-EEND method for various overlap situations, we prepared two training sets and five test sets, including simulated and real datasets.
The statistics of the training and test sets are listed in Table \ref{tab:set}.
The overlap ratio is computed as the ratio of the audio time during which two or more speakers are active, to the audio time during which one or more speakers are active.

Note that training data for the EEND method is different from those for the x-vector clustering-based method.
Whereas the x-vector clustering-based method uses single-speaker segments for training their x-vector neural network, the EEND method uses audio mixtures of multiple speakers. Such mixtures can be simulated infinitely with a combination of single-speaker segments. Moreover, the EEND model can be trained with not only simulated mixtures but real audio mixtures with speaker overlaps.

\subsubsection{Simulated mixtures}

%To verify the effectiveness of the SA-EEND method for controlled overlap situations, we conducted an evaluation with simulated mixtures.

Each mixture was simulated by Algorithm \ref{alg:mixture_simulation}.
Unlike the mixture simulation of source separation studies \cite{Hershey2016}, we consider a diarization-style mixture: each speech mixture should have dozens of utterances per speaker with reasonable silence intervals between utterances.
The silence intervals are controlled by the average interval of $\beta$. Larger values of $\beta$ generate speech with less overlap.
\begin{algorithm}[t]
	\SetAlgoLined
	\DontPrintSemicolon
	\caption{Mixture simulation.}
	\label{alg:mixture_simulation}
	\SetAlgoVlined
	\SetKwInOut{Input}{Input}
	\SetKw{In}{in}
	\Input{
	    {{$\mathcal{S,N,I,R}$} \tcp*{Set of speakers, noises, RIRs and SNRs}}\\
	    {{$\mathcal{U} = \{U_s\}_{s \in \mathcal{S}}$} \tcp*{Set of utterance lists}}
	    {{$N_\text{spk}$} \tcp*{\#speakers per mixture}}
	    {{$N_\text{umax},N_\text{umin}$} \tcp*{Max. and min. \#utterances per speaker}}
	    {{$\beta$} \tcp*{Average interval}}
	}
	\SetKwInOut{Output}{Output}
	\Output{$\mathbf{y}$\tcp*{Mixture}}
	\BlankLine
	Sample a set of $N_\text{spk}$ speakers $\mathcal{S'}$ from $\mathcal{S}$\\
		$\mathcal{X}\leftarrow\emptyset$\tcp*{Set of $N_\text{spk}$ speakers' signals}
		\ForAll{$s\in\mathcal{S'}$}{
			$\mathbf{x}_s\leftarrow\emptyset$\tcp*{Concatenated signal}
			Sample $\mathbf{i}$ from $\mathcal{I}$\tcp*{RIR}
			Sample $N_u$ from $\left\{N_\text{umin},\dots,N_\text{umax}\right\}$%\tcp*{\#utterances per speaker}
			
			\For{$u=1$ \rm{to} $N_u$}{
				Sample $\delta\sim\frac{1}{\beta}\exp\left(-\frac{\delta}{\beta}\right)$\tcp*{Interval}
				$\mathbf{x}_s\leftarrow\mathbf{x}_s\oplus\mathbf{0}^{\left(\delta\right)}\oplus U_s\left[u\right]\ast\mathbf{i}$
			}
			$\mathcal{X}.\mathsf{add}\left(\mathbf{x}_s\right)$\\
		}
		$L_\mathrm{max}=\max_{\mathbf{x}\in\mathcal{X}}\lvert\mathbf{x}\rvert$\\
		$\mathbf{y}\leftarrow\sum_{\mathbf{x}\in\mathcal{X}}\left(\mathbf{x}\oplus\mathbf{0}^{\left(L_\mathrm{max}-\lvert\mathbf{x}\rvert\right)}\right)$\\
		Sample $\mathbf{n}$ from $\mathcal{N}$\tcp*{Background noise}
		Sample $r$ from $\mathcal{R}$\tcp*{SNR}
		Determine a mixing scale $p$ from $r,\mathbf{y},$ and $\mathbf{n}$\\
		$\mathbf{n}'\leftarrow$ repeat $\mathbf{n}$ until reach the length of $\mathbf{y}$\\
		$\mathbf{y}\leftarrow\mathbf{y}+p\cdot\mathbf{n}'$\\
\end{algorithm}

The set of utterances used for the simulation was comprised of the Switchboard-2 (Phase I, II, III), Switchboard Cellular (Part 1, Part2), and NIST Speaker Recognition Evaluation datasets (2004, 2005, 2006, 2008). All recordings are telephone speech sampled at 8 kHz. There are 6,381 speakers in total. We split them into 5,743 speakers for the training set and 638 speakers for the test set.
Note that the set of utterances for the training set is identical to that of the Kaldi CALLHOME diarization v2 recipe \cite{Povey_ASRU2011}\footnote{\url{https://github.com/kaldi-asr/kaldi/tree/master/egs/callhome_diarization}}, making it fair comparison with the x-vector clustering-based method.

Since there are no time annotations in these corpora, we extracted utterances using speech activity detection (SAD) on the basis of time-delay neural networks and statistics pooling\footnote{The SAD model: \url{http://kaldi-asr.org/models/m4}}.

The set of background noises was from the MUSAN corpus \cite{Snyder2015}. We used 37 recordings that are annotated as ``background'' noises.
The set of 10,000 room impulse responses (RIRs) was from the Simulated Room Impulse Response Database used in \cite{Ko2017}.
%The total number of RIRs is 10,000.
The SNR values were sampled from 10, 15, and 20 dBs.
These sets of non-speech corpora are also used for training the x-vector and SAD models in the x-vector clustering-based method.

We generated two-speaker mixtures for each speaker with 10-20 utterances ($N_{\text{spk}} = 2, N_{\text{umin}}=10, N_{\text{umax}}=20$).
For the simulated training set, 100,000 mixtures were generated with $\beta=2$. For the simulated test set, 500 mixtures were generated with $\beta=2$, 3, and 5.
The overlap ratios of the simulated mixtures are ranging from 19.5 to 34.4\%.

\subsubsection{Real datasets}

We used real telephone speech recordings as the real training set. A set of 26,172 two-speaker recordings were extracted from the recordings of the Switchboard-2 (Phase I, II, III), Switchboard Cellular (Part 1, Part 2), and NIST Speaker Recognition Evaluation datasets.
The overlap ratio of the training data was 3.7\%, far less than that of the simulated mixtures.

We evaluated the proposed method on real telephone conversations in the CALLHOME dataset \cite{callhome}.
We randomly split the two-speaker recordings from the CALLHOME dataset into two subsets: an adaptation set of 155 recordings and a test set of 148 recordings.
The average overlap ratio of the test set was 13.0\%.

In addition, we conducted an evaluation on the dialogue part of the Corpus of Spontaneous Japanese (CSJ) \cite{Maekawa2003}. The CSJ contains 54 two-speaker dialogue recordings\footnote{We excluded four out of 58 recordings that contain speakers in the official speech recognition evaluation sets.}. They were recorded using headset microphones in separate soundproof rooms. The average overlap ratio of the CSJ test set was 20.1\%, larger than the CALLHOME test set.

\subsection{Model configuration}
\subsubsection{Clustering-based systems}
We compared the proposed method with two conventional clustering-based systems \cite{Sell2018dihard}: the i-vector system and x-vector system were created using the Kaldi CALLHOME diarization v1 and v2 recipes.
%As noted in Sec. \ref{sec:data}, we used the set of utterances and non-speech data that are the same as the simulated mixtures, for fair comparison with EEND methods.

These recipes use agglomerative hierarchical clustering (AHC) with the probabilistic linear discriminant analysis (PLDA) scoring scheme. The number of clusters was fixed to 2. Though the original recipes use oracle speech/non-speech marks, we used the SAD model with the same configuration as described in Sec. \ref{sec:data}.

\subsubsection{BLSTM-based EEND system}

We configured a BLSTM-based EEND method (BLSTM-EEND), as described in \cite{Fujita2019E2EDiarization}. 
The input features were 23-dimensional log-Mel-filterbanks with a 25-ms frame length and 10-ms frame shift. Each feature was concatenated with those from the previous seven frames and subsequent seven frames. To deal with a long audio sequence in our neural networks, we subsampled the concatenated features by a factor of ten. Consequently, a $(23\times 15)$-dimensional input feature was fed into the neural network every 100 ms.

We used a five-layer BLSTM with 256 hidden units in each layer. The second layer of the BLSTM outputs was used to form a 256-dimensional embedding; we then calculated the deep clustering loss in this embedding to discriminate different speakers. We used the Adam \cite{Kingma2014} optimizer with a learning rate of $10^{-3}$. The batch size was 10. The number of training epochs was 20.

Because the output of the neural network is the probability of speech activity for each speaker, a threshold is required to obtain the decision of speech activity for each frame. We set the threshold to 0.5. Furthermore, we applied 11-frame median filtering to prevent production of unreasonably short segments.

For domain adaptation, the neural network was retrained using the CALLHOME adaptation set.
we used the Adam optimizer with a learning rate of $10^{-6}$ and ran 5 epochs.
For the postprocessing, we adjusted the threshold to 0.6 so that the DER of the adaptation set has the minimum value.

\subsubsection{Self-attentive EEND system}

Here, we used the same input features as were input to the BLSTM-EEND system. Note that the sequence length at the training stage was limited to 500 (50 seconds in audio time) because our system uses more memory than the BLSTM-based network does. Therefore, we split the input audio recordings into non-overlapping 50-second segments. At the inference stage, we used the entire sequence for each recording.

We used two encoder blocks with 256 attention units containing four heads ($P=2$, $D=256$, $H=4$). We used 1024 internal units in a position-wise feed-forward layer ($d_\mathrm{ff}=1024)$.
We used the Adam optimizer with the learning rate scheduler introduced in \cite{Vaswani2017}. The number of warm-up steps used in the learning rate scheduler was 25,000. The batch size was 64. The number of training epochs was 100.
After 100 epochs, we used an averaged model obtained by averaging the model parameters of the last 10 epochs.
As with the BLSTM-EEND system, we applied 11-frame median filtering.

For domain adaptation, the averaged model was retrained using the CALLHOME adaptation set.
We used the Adam optimizer with a learning rate of $10^{-5}$ and ran 100 epochs.
After 100 epochs, we used an averaged model obtained
by averaging the model parameters of the last 10 epochs.

\subsection{Performance metric}

We evaluated the systems with the diarization error rate (DER) \cite{NISTRT09}. Note that the DERs reported in many prior studies did not include misses or false alarm errors due to their using oracle speech/non-speech labels. Overlapping speech segments had also been excluded from the evaluation. For our DER computation, we evaluated all of the errors, including overlapping speech segments, because the proposed method includes both the speech activity detection and overlapping speech detection functionality. As is done typically, we used a collar tolerance of 250 ms at the start and end of each segment.

\section{Results}
\label{sec:results}

%%% Big table with sim. variations
\begin{table}[t]
\caption{DERs (\%) on various test sets. For EEND systems, the CALLHOME (CH) results are obtained with domain adaptation.
}
\label{tab:overlap}
\centering
\begin{tabular}{l|ccc|cc} \hline
 & \multicolumn{3}{c|}{Simulated} & \multicolumn{2}{c}{Real} \\
 & $\beta=2$ & $\beta=3$ & $\beta=5$ & CH & CSJ \\ \hline 
Clustering-based & & & & &\\
\: i-vector & 33.74 & 30.93 & 25.96 & 12.10 & 27.99 \\
\: x-vector &	28.77 & 24.46 & 19.78 & 11.53 & 22.96 \\ \hline
BLSTM-EEND & & & & & \\
\: trained with sim. & 12.28 & 14.36 & 19.69 & 26.03 & 39.33 \\
\: trained with real & 36.23 & 37.78 & 40.34 & 23.07 & 25.37 \\ \hline
SA-EEND & & & & &\\
\: trained with sim.  & {\bf 7.91} & {\bf 8.51} & {\bf 9.51} & 13.66 & 22.31 \\
\: trained with real  & 32.72 & 33.84 & 36.78 & {\bf 10.76} & \bf 20.50 \\
\hline
\end{tabular}
\end{table}

\begin{table}[t]
\caption{DERs (\%) on the CALLHOME with and without domain adaptation.
}
\label{tab:adapt}
\centering
\begin{tabular}{l|cc} \hline
 & w/o adaptation & with adaptatation \\ \hline
x-vector clustering & 11.53 & N/A \\ \hline
BLSTM-EEND \\
\: trained with sim. & 43.84 & 26.03 \\
\: trained with real & 31.01 & 23.07 \\ \hline
SA-EEND \\
\: trained with sim. & 17.42 & 13.66 \\
\: trained with real & 12.66 & \bf{10.76} \\ \hline
\end{tabular}
\end{table}

\begin{table}[t]
\caption{Detailed DERs (\%) evaluated on the CALLHOME. DER is composed of Misses (MI), False alarms (FA), and Confusion errors (CF). The SAD errors are composed of Misses (MI) and False alarms (FA) errors.}
\label{tab:detail}
\centering
\begin{tabular}{l|c|ccc|cc} \hline
    & & \multicolumn{3}{c|}{DER breakdown} & \multicolumn{2}{c}{SAD errors} \\
Method & DER & MI & FA & CF & MI & FA \\ \hline
i-vector & 12.10 & 7.74 & 0.54 & 3.82 & 1.4 & 0.5 \\
x-vector &	11.53 & 7.74 & 0.54 & 3.25 & 1.4 & 0.5 \\ \hline
\begin{tabular}[b]{@{}c@{}}SA-EEND\\\quad no-adapt\end{tabular} &	12.66 & 7.42 & 3.93 &  1.31 & 3.3 & 0.6 \\
\quad adapted  & {\bf 10.76} & 6.68 & 2.40 &  1.68 & 2.3 & 0.5 \\ \hline
\end{tabular}
\end{table}

\begin{figure*}[t]
\begin{center}
    \includegraphics[width=\linewidth]{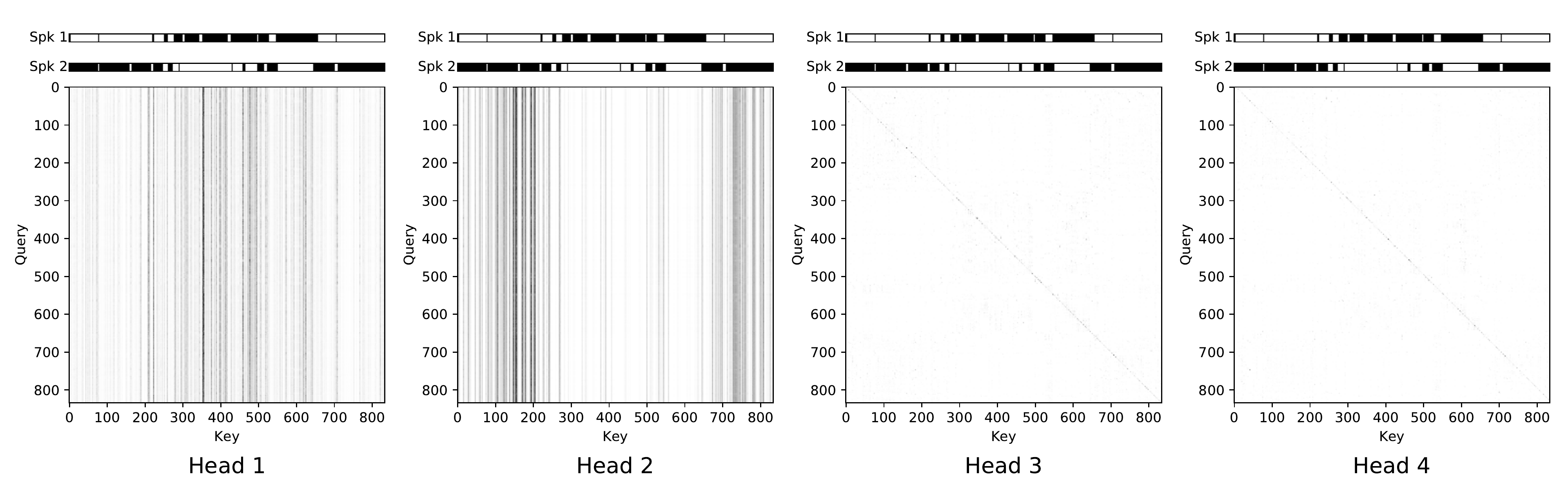}
\end{center}
\caption{Attention weight matrices at the second encoder block. The input was the CALLHOME test set (recording id: iagk). The model was trained with the real training set followed by domain adaptation. The top two rows show the reference speech activity of two speakers.}
\label{fig:visualize}
\end{figure*}

%To investigate the robustness to variable conditions, we evaluated the diarization systems on various test sets; (i) speaker-overlapping simulated mixtures with three levels of overlap ratios, (ii) a real telephone dataset with small speaker overlaps (CALLHOME), and (iii) a real dialogue dataset with relatively large speaker overlaps (CSJ). The results are shown in Table \ref{tab:overlap}.

\subsection{Evaluation on simulated mixtures}

DERs on various test sets are shown in Table \ref{tab:overlap}.
The clustering-based systems performed poorly on heavily overlapped simulated mixtures.
This result is within our expectations, because the clustering-based systems did not consider speaker overlaps; there are more misses when the overlap ratio is high.

The BLSTM-EEND system trained with the simulated training set showed a significant DER reduction compared with the clustering-based systems on the simulated mixtures.
Among the differing overlap ratios, it showed the best performance on the highest overlap ratio condition ($\beta=2$).
The BLSTM-EEND system worked well on the overlapping condition matched with training data.

The proposed system, SA-EEND, trained with the simulated training set had significantly fewer DERs compared with the BLSTM-EEND system on every test set.
As well as the BLSTM-EEND system, it showed the best performance on the highest overlap ratio condition ($\beta=2$).
However, the DER degradation on the less overlapping conditions was smaller than that of the BLSTM-EEND system, which indicated that the self-attention blocks improved robustness to variable overlapping conditions.

\subsection{Evaluation on real test sets}

In contrast to the good performance on the simulated mixtures, the BLSTM-EEND system had inferior DERs to those of the clustering-based systems evaluated on the real test sets.
Although the BLSTM-EEND system showed performance improvements when the training data were switched from simulated to real data, its DERs were still higher than those of the clustering-based systems. 

The proposed system, SA-EEND, trained with the simulated training set showed remarkable improvements on real datasets of the CALLHOME and CSJ, which indicates the strong generalization capability of the self-attention blocks.
For the CSJ, even without domain adaptation, the proposed system performed better than the x-vector clustering-based method.

The SA-EEND system trained with the real training set performed the best on the real test sets, however, it had poor DERs on the simulated mixtures. We expected that the result was due to the small number of mixtures and low overlap ratio of the real training set. It would be much improved by feeding more real data with more speaker overlaps, or by combining with simulated training data.

\subsection{Effect of domain adaptation}

The EEND models trained with simulated training set were overfitted to the specific overlap ratio of the training set.
We expected that the overfitting would be mitigated by using domain adaptation.
DERs on the CALLHOME with and without domain adaptation are shown in Table \ref{tab:adapt}.
As expected, the domain adaptation significantly reduced the DER; our system thus achieved even better results than those of the x-vector-based system.

A detailed DER comparison on the CALLHOME test set is shown in Table \ref{tab:detail}. The clustering-based systems had few SAD errors thanks to the robust SAD model trained with various noise-augmented data. However, there were numerous misses and confusion errors due to its lack of handling speaker overlaps. Compared with clustering-based systems, the proposed method produced significantly fewer confusion and miss errors.
The domain adaptation reduced all error types except confusion errors.

\subsection{Visualization of self-attention}

To analyze the behavior of the self-attention mechanism in our diarization system, Fig. \ref{fig:visualize} visualizes the attention weight matrix at the second encoder block, corresponding to $\mathbf{\hat{A}}^{(p=2)}_h$ in Eq. \ref{eq:softmax}.
Here, head 1 and head 2 have vertical lines at different positions.
The vertical lines correspond to each speaker's activity.
The attention weight matrix with these vertical lines transformed the input features into the weighted mean of the same speaker frames. These heads actually captured the global speaker characteristics by computing the similarity between distant frames.
%The input features were transformed via these matrices into the weighted sum of the same speaker frames.
%The vertical lines were observed at distant positions.
%This result indicates the self-attention mechanism captures the similarity between such distant frames, which leads to a reduction in DER.
Interestingly, heads 3 and 4 look like identity matrices, which results in position-independent linear transforms. These heads are considered to work for speech/non-speech detection.
We conclude that the multi-head self-attention mechanism captures global speaker characteristics in addition to local speech activity dynamics, which leads to a reduction in DER.
Experiments on various combinations of the number of heads and the number of speakers would be an interesting future work.

\section{Conclusion}
\label{sec:print}

We incorporated a self-attention mechanism in the end-to-end neural diarization model. We evaluated our model on simulated mixtures and two real datasets. Experimental results showed that the self-attention mechanism significantly reduced DERs and showed higher generalization quality compared with a BLSTM-based neural diarization system. The self-attention based systems even outperformed x-vector clustering-based systems.
We also showed that the self-attention blocks actually captured global speaker characteristics by visualizing the latent representation.
%We also showed that the similarity between distant frames is actually captured by the self-attention blocks by visualizing the latent representation. 

%\section{Acknowledgements}

%\bibliographystyle{IEEEbib}
\bibliographystyle{IEEEbib_abbrev}
\bibliography{refs,refs_regular}

\end{document}